\documentclass[twocolumn,noshowpacs,preprintnumbers,amsmath,amssymb,aps]{revtex4-2}
\usepackage{graphicx}
\usepackage{dcolumn}
\usepackage{bm}

\begin{document}

\title{Demonstration of a superconducting diode-with-memory, operational at zero magnetic field with switchable nonreciprocity. }

\author{Taras Golod}
\author{Vladimir M. Krasnov}
\email{Vladimir.Krasnov@fysik.su.se}

\affiliation{Department of Physics, Stockholm University,
AlbaNova University Center, SE-10691 Stockholm, Sweden.}


\begin{abstract}
Diode is one of basic electronic components. It has a nonreciprocal current response, associated with a broken space/time reversal symmetry. Here we demonstrate prototypes of superconducting diodes operational at zero magnetic field. They are based on conventional niobium planar Josephson junctions, in which space/time symmetry is broken by a combination of self-field effect from nonuniform bias and stray fields from a trapped Abrikosov vortex. We demonstrate that nonreciprocity of critical current 
in such diodes can reach an order of magnitude and rectification efficiency can exceed 70$\%$. Furthermore, we can easily change the diode polarity and switch nonreciprocity on/off by changing bias configuration and by trapping/removing a vortex. This facilitates memory functionality. We argue that such diode-with-memory can be used for a future generation of in-memory superconducting computers.

\end{abstract}

\maketitle

\section{Introduction}

Large computation facilities, such as big data centers 
and supercomputers have become major energy consumers with a power budget often in excess of 100 MW. 
It has been argues that a small fraction of this power 
would be sufficient for cooling down the facility to cryogenic temperatures, suitable for operation of superconductors (SC) \cite{Holmes_2013}. 
SC electronics would not only enable effective utilization of energy by removing resistive losses, it could also greatly enhance the operation speed. Since there is no resistance, $R=0$, the $RC$ time constant is no longer a limiting factor. The ultimate operation frequency is determined by the SC energy gap. For many SCs it is in the THz range \cite{Borodianskyi_2017}. This enables clock frequencies several orders of magnitude higher than for modern semiconducting electronics. 
Such perspectives has lead to a renewed interest in development of a digital SC computer \cite{Holmes_2013,Golod_2015,Soloviev_2017,Herr_2018,Tolpygo_2019,Semenov_2019}.

Diode is one of the primary electronic components. Its nonreciprocal current-voltage ($I$-$V$) characteristics allows rectification of alternating currents, which is necessary for signal processing and ac-dc conversion. Diodes can be also used as building blocks for Boolean logics in digital computation. SC diodes should have strongly asymmetric critical currents, $|I_{c+}| \ne |I_{c-}|$. It is well known that nonreciprocity may appear in spatially asymmetric SC devices \cite{Likharev_1981,Barone_1982}. SC diodes, based on spatially nonuniform Josephson junctions (JJs), were demonstrated long time ago \cite{Krasnov_1997}. Also SC ratchets \cite{Hanggi_2009}, rectifying motion of either Josephson \cite{Falo_2002,Shalom_2005,Beck_2005,Wang_2009} or Abrikosov \cite{Janko_1999,Villegas_2003,Tonomura_2005,Moshchalkov_2006,Harrington_2009,Silhanek_2013,Lustikova_2018,Kwok_2021} vortices, were intensively studied. However, such spatially asymmetric devices operate only at finite magnetic fields, while computer components should work at zero field. Nonreciprocity at $H=0$ is prohibited by the time-reversal symmetry, which requires invariance of electromagnetic characteristics upon simultaneous flipping of current and magnetic field \cite{Krasnov_1997,Albert}. Therefore, zero-field SC diode requires breaking of both space and time reversal symmetry.

Recently it was shown that nonreciprocity can be induced in noncentrosymmetric SC by spin-orbit interaction (SOI) \cite{Nagaosa_2017,Nagaosa_2018,Qin_2017,Tokura_2018}. This renewed search for diode effects in noncentrosymmetric SC  \cite{Hu_2007,Nagaosa_2017,Qin_2017,Zhang_2020} and heterostructures \cite{Ando_2020,Strunk_2022,Ali_2021}.
SOI can induce asymmetry of either resistance in the fluctuation region near $T_c$  \cite{Nagaosa_2017,Nagaosa_2018,Qin_2017,Tokura_2018,Yasuda_2019,Zhang_2020}, or supercurrent at low $T$ \cite{Ando_2020,Strunk_2022,Pal_2021,Lin_2021,Diez_2021,Bauriedl_2021,Shin_2021}.
However, SOI-based diodes require significant magnetic field. In several works zero-field SC diode operation was reported \cite{Ali_2021,Lin_2021}, involving additional nontrivial effects. In this respect, nonreciprocity can be a tool for investigation of  unconventional SC  \cite{Yasuda_2019,Pal_2021,Lin_2021,Diez_2021,Bauriedl_2021,Shin_2021}.

In this work we demonstrate prototypes of SC diodes with a large and switchable nonreciprocity of supercurrent at zero magnetic field. They are made of a conventional Nb SC and contain cross-like planar Josephson junctions with additional electrodes and an artificial vortex trap. Nonreciprocity is induced by a combination of self-field effect from asymmetric bias and stray fields from trapped Abrikosov vortex (AV).
We demonstrate that the ratio, $|I_{c+}/I_{c-}|$, of such diodes can reach an order of magnitude and rectification efficiency can exceed 70$\%$.
Furthermore, we can switch nonreciprocity on and off, as well as change diode polarity in one and the same device. This is achieved by trapping/removing either a vortex, or an antivortex, and/or by changing bias configuration. This facilitates memory functionality. We argue that such diode-with-memory can be used for a new generation of superconducting in-memory computers.

\section{The concept}

\begin{figure*}[t]
    \centering
    \includegraphics[width=0.99\textwidth]{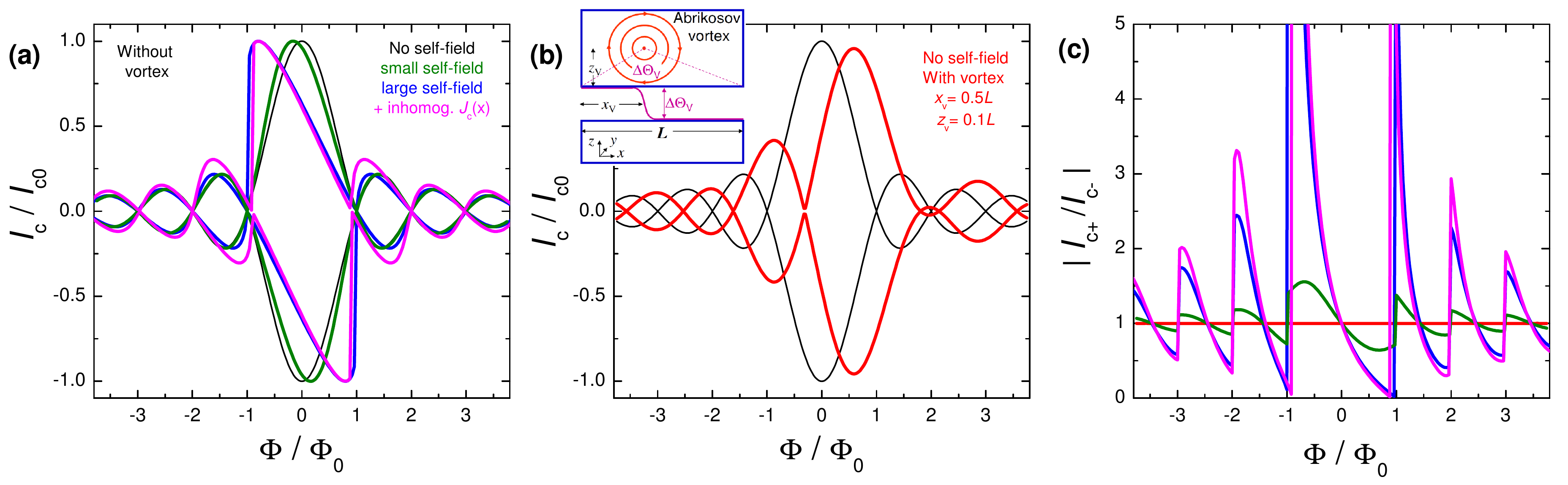}
    \caption{ {\bf Numerical modelling of the two diode ingredients: (a) self-field induced by nonuniformity and (b) persistent stray fields from a trapped Abrikosov vortex.} (a) Simulated $I_c(\Phi)$ modulation: black - for a uniform JJ with constant critical current density, $J_c$, and bias, $I_b$; olive - for $J_c=const$ with a slightly nonuniform bias, $I_b(x)$, leading to appearance of self-field; blue - for $J_c=const$ with a strongly nonuniform bias, $I_b(x)$; magenta - the same as for blue with added V-shape spatial inhomogeneity of $J_c(x)$. (b) Simulated $I_c(\Phi)$ for a uniform JJ with a uniform bias without (black) and with a trapped antivortex (red) at $x_v=L/2$ and $z_v=0.1~L$. Inset shows a sketch of vortex-junction configuration. (c) Nonreciprocity,  $|I_{c+}(\Phi)/I_{c-}(\Phi)|$, for the cases from (a) and (b) in the same color palette. It is seen that nonuniformity induces nonreciprocity, but only at a finite field, $H \propto \Phi \ne 0$. Vortex stray fields shift and distort $I_c(\Phi)$, but do not induce nonreciprocity.
    }
    \label{fig:fig1}
\end{figure*}

We consider the simplest case of a short JJ with the length $L< 4\lambda_J$, where $\lambda_J$ is the Josephson penetration depth. This allows neglecting of complex phenomena associated with screening effects and Josephson vortices \cite{Barone_1982,Krasnov_1997,Krasnov_2020}. 
Realization of zero-field SC diode requires breaking of space/time symmetry. 
Time-reversal leads to inversion of transport currents and magnetic fields 
generated by these currents. The role of external field, $H$, is somewhat more tricky \cite{Albert}. However, since it induces spatial phase gradient in a JJ, it is connected with the spatial symmetry \cite{Krasnov_2020}.

Our concept has two simple ingredients: (i) Utilization of nonuniform bias for achieving nonreciprocity {\em at finite fields} \cite{Krasnov_1997}; and (ii) Shifting it to {\em zero field} by persistent stray fields from trapped AV \cite{Golod_2010,Golod_2019b,Krasnov_2020}. These effects are summarized in Figs. \ref{fig:fig1} (a) and (b). Here black lines represent the conventional Fraunhofer modulation of the critical current versus magnetic flux, $I_c(\Phi)$, for a uniform JJ without a vortex. In this case there are both time-reversal, $I_{c+}(H)=|I_{c-}(H)|$, and space-reversal, $I_{c\pm}(H)=I_{c\pm}(-H)$, symmetries.

Nonuniformity of junction characteristics breaks spatial symmetry. Most common are nonuniform critical current density,  $J_c(x)$, and bias current distribution, $I_b(x)$. Both lead to appearance of self-field effect \cite{Likharev_1981,Barone_1982,Krasnov_1997}: nonuniformly distributed current generates magnetic field component parallel to $H$. Olive and blue lines in Fig. \ref{fig:fig1} (a) represent calculated $I_c(\Phi)$ patterns (see Methods) for small and large self-field effects in a JJ with nonuniform bias, $I_b(x)\ne const$, but uniform $J_c(x)=const$. Magenta lines are calculated for the same $I_b(x)\ne const$ as for the blue curves, with additional nonuniformity of $J_c(x)\ne const$ . As discussed in Ref. \cite{Krasnov_1997}, asymmetric nonuniformities of both $J_c(x)$ and $I_b(x)$ tilt $I_c(H)$ patterns and lead to appearance of nonreciprocity,  $I_{c+}(H) \ne |I_{c-}(H)|$, at finite $H$. However, $I_c(H)$ remain centrosymmetric, $I_{c+}(H) = -I_{c-}(-H)$, for any nonuniformity. This is the consequence of space/time symmetry: simultaneous flipping of $I$ and $H$ is equivalent to looking at the same JJ from the back side and, therefore, should lead to the identical observation \cite{Krasnov_1997}.

Red lines in Fig. \ref{fig:fig1} (b) represent $I_c(\Phi)$ modulation in a uniform junction with a trapped AV, placed symmetrically in the middle of the electrode, $x_v=L/2$, at a distance $z_v=0.1~L$ from the JJ, see the sketch in the inset. Stray fields from AV both distort and shift the $I_c(H)$ pattern \cite{Golod_2010,Golod_2019b,Golod_2021}. This breaks space-reversal, $I_{c\pm}(H) \ne I_{c\pm}(-H)$, but preserves time-reversal, $I_{c+}(H)=|I_{c-}(H)|$, symmetry. Note that for short JJs this symmetry is preserved even for asymmetric vortex locations, $x_v\ne L/2$, but for long JJs all types of symmetries are removed due to appearance of Josephson vortices \cite{Krasnov_2020}.

In Fig. \ref{fig:fig1} (c) we plot nonreciprocity, $|I_{c+}(\Phi)/I_{c-}(\Phi)|$, for the curves from panels (a) and (b). It is seen that uniform JJs without (black) or with (red) a vortex are reciprocal, $|I_{c+}/I_{c-}|=1$. Nonuniform JJs exhibit nonreciprocity, which grows with increasing inhomogeneity of $J_c(x)$ and $I_b(x)$. Such Josephson diodes were studied earlier \cite{Krasnov_1997}. Their nonreciprocity could be very high: the largest peak at $\Phi/\Phi_0\simeq -1$ for the magenta curve in Fig. \ref{fig:fig1} (c) reaches almost two orders of magnitude. However, due to centrosymmetric $I_{c}(H)$ there is no effect at $H=0$.

\begin{figure}[t]
    \centering
    \includegraphics[width=0.47\textwidth]{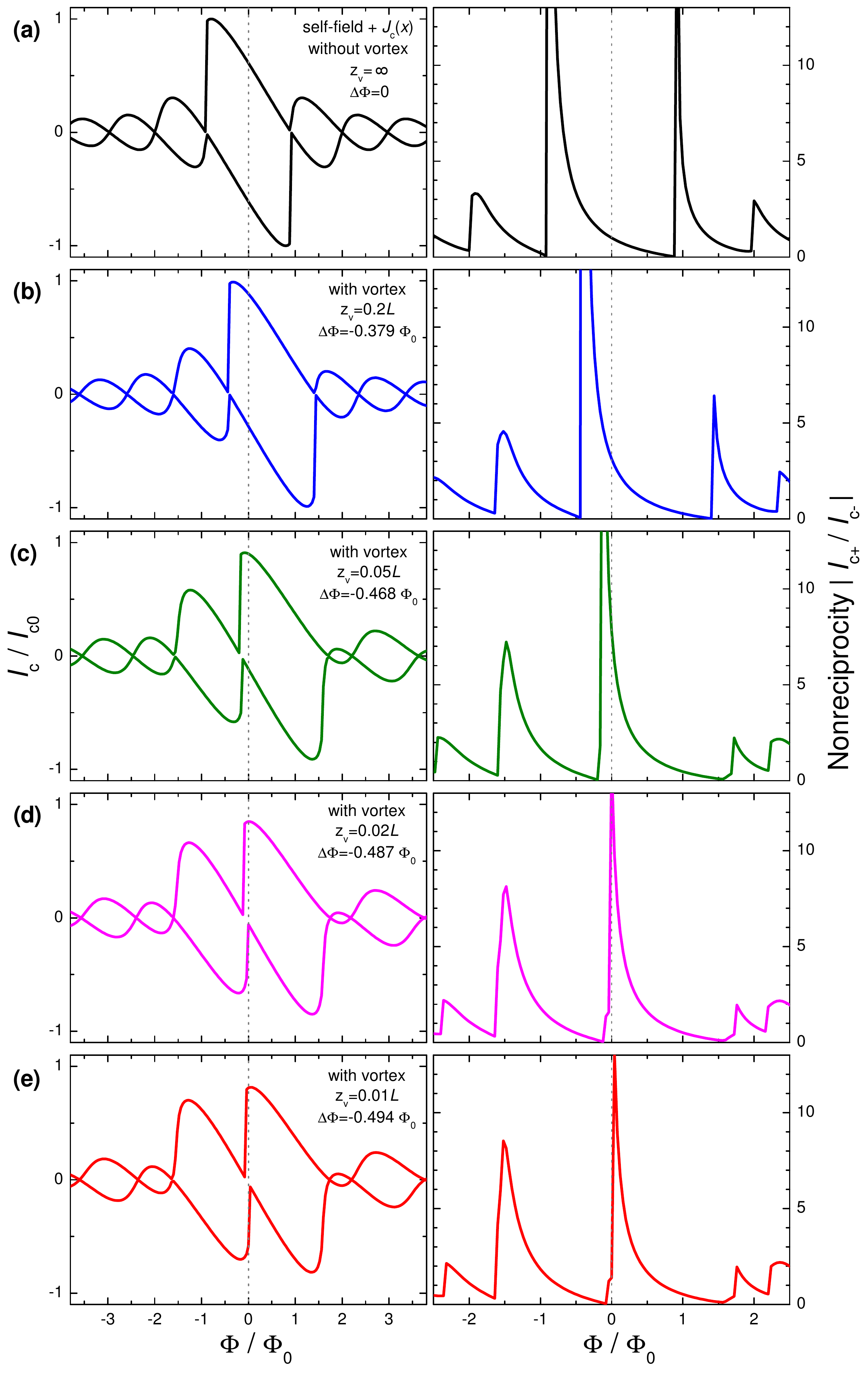}
    \caption{ {\bf Numerical modelling of combined effects of nonuniformity and stray magnetic fields from a trapped vortex.} Left panels show simulated $I_c(H)$ patterns for a nonuniform JJ (the same as shown by magneta curves in Fig. \ref{fig:fig1}) upon approaching an Abrikosov vortex along the middle line, $x_v=L/2$, from (a) $z_v=\infty$ to (e) $z_v=0.01~L$. Right panels represent corresponding nonreciprocities, $|I_{c+}(H)/I_{c-}(H)|$. It is seen that upon approaching the vortex to the junction, growing stray fields progressively shifts nonreciprocal peaks. At a certain distance (d) the main peak passes through $H=0$. This is the optimal geometrical configuration for a zero-field diode.
    }
    \label{fig:fig2}
\end{figure}

For shifting nonreciprocity to $H=0$, we utilize persistent stray fields from a trapped AV. As shown in Refs. \cite{Golod_2019b,Golod_2021}, flux offset introduced by AV in planar JJs is determined by the polar angle $\Theta_v$ of the vortex within the junction, see the sketch in Fig. 1 (b). It depends on the vortex location ($x_v,~z_v)$,
which is the tunning geometrical factor for the vortex-induced flux offset \cite{Golod_2019b,Golod_2021}. 
In Figure \ref{fig:fig2} we show variation of $I_c(\Phi)$ modulation (left) and nonreciprocity, $|I_{c+}(\Phi)/I_{c-}(\Phi)|$, (right panels) upon approaching AV towards the junction from (a) $z_v=\infty$ to (e) $z_v=0.01~L$ along the middle line $x_v=L/2$. Simulations are done for the same nonuniform JJ depicted by the magenta line in Fig. \ref{fig:fig1}.  It is seen that with approaching the vortex to the JJ, the central nonreciprocal peak moves gradually from $\Phi/\Phi_0 \simeq -1$ towards 0 without significant reduction of the amplitude. At (d) $z_v=0.05~L$ it passes through $\Phi = 0$. This is the optimal geometrical configuration for zero-field diode operation.

\section{Experimental results}

\begin{figure*}[t]
    \centering
    \includegraphics[width=0.9\textwidth]{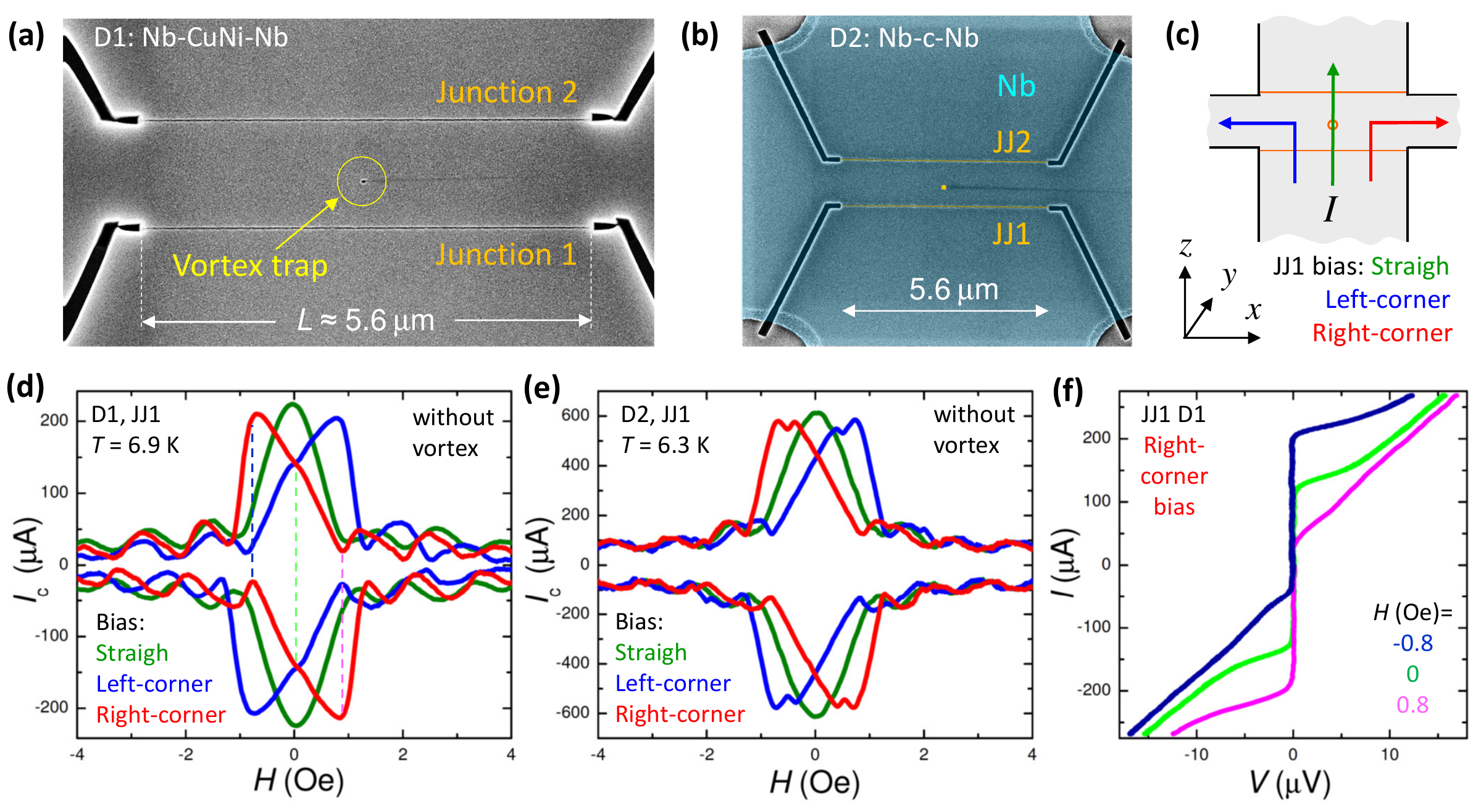}
    \caption{ {\bf Device  characterisation in the vortex-free case.} (a) and (b) SEM images of two studied devices: (a) D1 with Nb-CuNi-Nb junctions and (b) D2 with constriction-like Nb-c-Nb variable thickness bridges (false color). They have similar geometries and contain two planar junctions and a vortex trap. As seen from (b), devices have cross-like geometry with four electrodes (top, bottom, left, right). This allows controllable variation of self-field effect by changing bias configurations, as sketched in (c). Field is applied perpendicular to the film, in the $y$-direction, ($x,y,z$) is the right-handed coordinate system. (d) and (e) Measured $I_c(H)$ patterns for junctions 1 on (d) D1 and (e) D2 devices for three bias configurations. Straight bias (olive lines) does not induce self-field and leads to a regular Fraunhofer modulation without nonreciprocity. Left (blue) and right (red) corner bias induces self-fileds of opposite signs, causing profound tilting of $I_c(H)$ patterns in opposite directions. (f) $I$-$V$ curves for JJ1 on D1 at three magnetic fields marked by dashed lines in (d). A profound nonreciprocity with a factor $\sim 10$ difference between $I_{c+}$ and $|I_{c-}|$ can be seen.
    }
    \label{fig:fig3}
\end{figure*}

Figures \ref{fig:fig3} (a) and (b) show scanning electron microscope (SEM) images of two studied devices, D1 and D2. They have similar geometries. Each contains two planar JJs with $L=5.6~\mu$m, seen as horizontal lines, and a vortex trap - a hole with diameter $\sim 50$ nm, placed at $x_v\simeq L/2$ and $z_v \simeq 0.1 ~L$ from JJ1. D1 is made from a Nb(70 nm)/CuNi(50 nm) bilayer with superparamagnetic CuNi, while D2 is made from a single Nb film (70 nm). Therefore, D1 has proximity-coupled Nb-CuNi-Nb JJs and D2 contains variable thickness type constriction JJs, Nb-c-Nb. Both devices behave in a similar way, but Nb-c-Nb JJs have much larger $I_c R_n$, which can exceed 1 mV at low $T$ \cite{Grebenchuk_2022}. This increases both the readout voltage and the upper operation frequency, $f_c=I_cR_n/\Phi_0$, which is advantageous for electronic applications. Details of junction fabrication, characterization and experimental setup can be found in Methods, Supplementary information \cite{Supplem} and Refs. \cite{Golod_2010,Golod_2015,KrasnovSFS_2005,Golod_2019a,Golod_2019b,Boris_2013,Grebenchuk_2022}.
Magnetic field is applied perpendicular to Nb film (in the $y$-direction).

As can be seen from Fig. \ref{fig:fig3} (b), studied devices have a cross-like geometry with four electrodes (left, right, top, bottom). This allows controllable introduction of asymmetric bias \cite{Golod_2019a}. In Fig. \ref{fig:fig3} (c) we sketch three bias configurations for JJ1. Straight (bottom-to-top) bias does not generate $H_y$ field component, while bias over right/left corners does induce positive/negative self-field $H_y$ in the junction. This strongly affects junction characteristics and facilitates tunable introduction of spatial asymmetry.

Figs. \ref{fig:fig3} (d) and (e) show $I_c(H)$ patterns of JJ1 in the vortex-free case, measured using the three bias configurations for (d) D1 and (e) D2. Straight bias (olive curves) leads to  regular Fraunhofer modulation. However, right (red) and  left (blue) corner biases tilt $I_c(H)$ patterns in opposite directions due to appearance of self-fields \cite{Krasnov_1997}. Fig. \ref{fig:fig3} (f) shows the $I$-$V$ characteristics at three fields indicated by dashed lines in Fig. \ref{fig:fig3} (d). At $H=0$ (green) the $I$-$V$ is symmetric $I_{c+}= |I_{c-}|$, however, at $H\simeq \pm 0.8$ Oe, profound nonreciprocities appear, reaching an order of magnitude of either sign.
This demonstrates that the cross-like geometry allows simple and controllable introduction of spatial (bias) asymmetry and associated nonreciprocity at $H\ne 0$.

\begin{figure*}[t]
    \centering
    \includegraphics[width=0.9\textwidth]{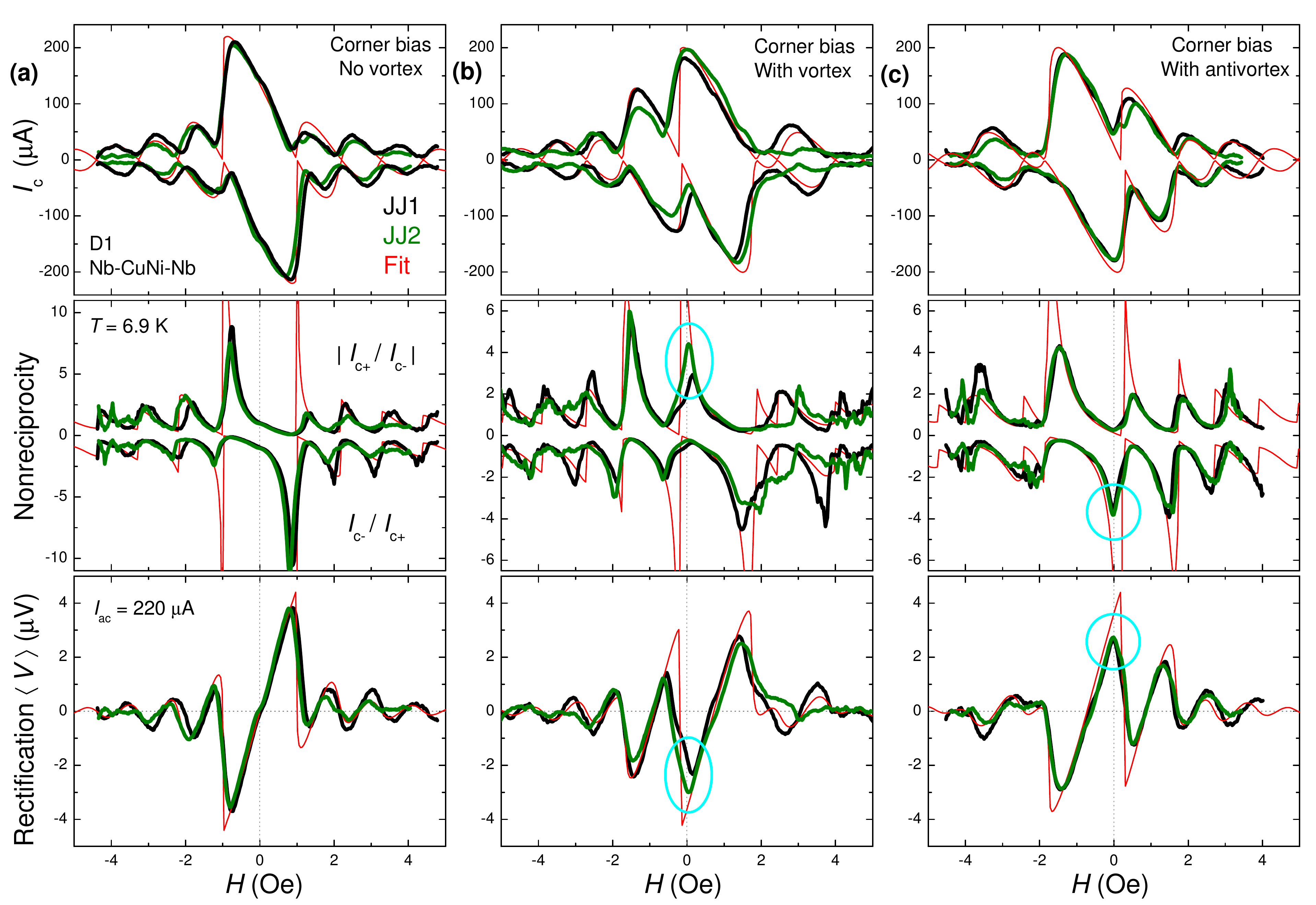}
    \caption{ {\bf Diode operation with the right-corner bias.} (a) Without a vortex, (b) with a trapped vortex and (c) with an antivortex. Top panels show $I_c(H)$. Middle panels show nonreciprocities $|I_{c+}/I_{c-}|$, upper curves, and $I_{c-}/I_{c+}$, lower curves. Bottom panels show rectified dc-voltage calculated for harmonic ac-bias with $I_{ac}=220~\mu$A$~\simeq I_{c0}$. Black and olive lines represent data for junctions 1 and 2 on D1, red lines are numerical fits. Appearance of profound nonreciprocity and rectification at $H=0$ is marked by cyan circles in (b) and (c). All measurements are performed at $T\simeq 6.9$ K.
    }
    \label{fig:fig4}
\end{figure*}

Fig. \ref{fig:fig4} summarizes diode performance for D1 with the right-corner bias (a) without AV, and (b) with a trapped vortex, or (c) antivortex. AV is controllably introduced and removed by short current pulses \cite{Golod_2015,Golod_2021}, as described in the Supplementary \cite{Supplem}. Top panels show $I_c(H)$ modulations for JJ1 (black) and JJ2 (olive). Red lines represent numerical fits (see Methods). In Fig. 4 (a) it is identical with the magenta curve in Fig. \ref{fig:fig1} (a). 
Fits in panels (b) and (c) are made for the actual geometry of the vortex trap 
and with AV-induced flux $\delta \Phi/\Phi_0 = \pm 0.47$ as a fitting parameter. It is consistent with $\Theta_{v1}/2\pi \simeq 0.44$ for JJ1 \cite{Supplem}.

Middle panels in Fig. \ref{fig:fig4} represent main experimental results of this work: nonreciprocities $|I_{c+}/I_{c-}|$ and $I_{c-}/I_{c+}$ for both JJs on D1. It can be seen that without AV, $|I_{c-}/I_{c+}|$ at $H\simeq 0.8$ Oe exceeds an order of magnitude, while it is absent at $H=0$. 
Introduction of AV shifts the maxima so that a significant nonreciprocity occurs at zero field, as indicated by cyan ovals in (b) and (c). The maximum for JJ2 in (b) exceeds a factor four. For JJ1 it is slightly offset (by $\sim 0.1$ Oe) but still exceeds a factor two at $H=0$. In (c) nonreciprocity at $H=0$ is more than three for both JJs.
Note, that diode polarity is opposite for vortex, $I_{c+} >|I_{c-}|$, and antivortex, $I_{c+} < |I_{c-}|$. Moreover, in our cross-like devices, the polarity can also be flipped by changing bias configuration. In Fig. \ref{fig:fig4} we use the right-corner bias. If we change to the left-corner bias, both self-field and polarity change sign, as demonstrated in Figs. \ref{fig:fig3} (d) and (e). For left-corner bias the polarity of vortex and antivortex states flips so that $|I_{c-}|>I_{c+}$ for vortex and $I_{c+}>|I_{c-}|$ for antivortex. All mentioned states are persistent and are achievable in one and the same device. Therefore, our diode is {\em switchable}. This enables memory functionality \cite{Golod_2015,Supplem} with three distinct states at $H=0$: a reciprocal state ``0" without AV, Fig. \ref{fig:fig4} (a), and states ``+1" and ``-1" with positive and negative polarities, shown in Figs. \ref{fig:fig4} (b) and (c). Such reconfigurability is a unique property of vortex-based devices \cite{Golod_2021}.

\begin{figure*}[t]
    \centering
    \includegraphics[width=0.99\textwidth]{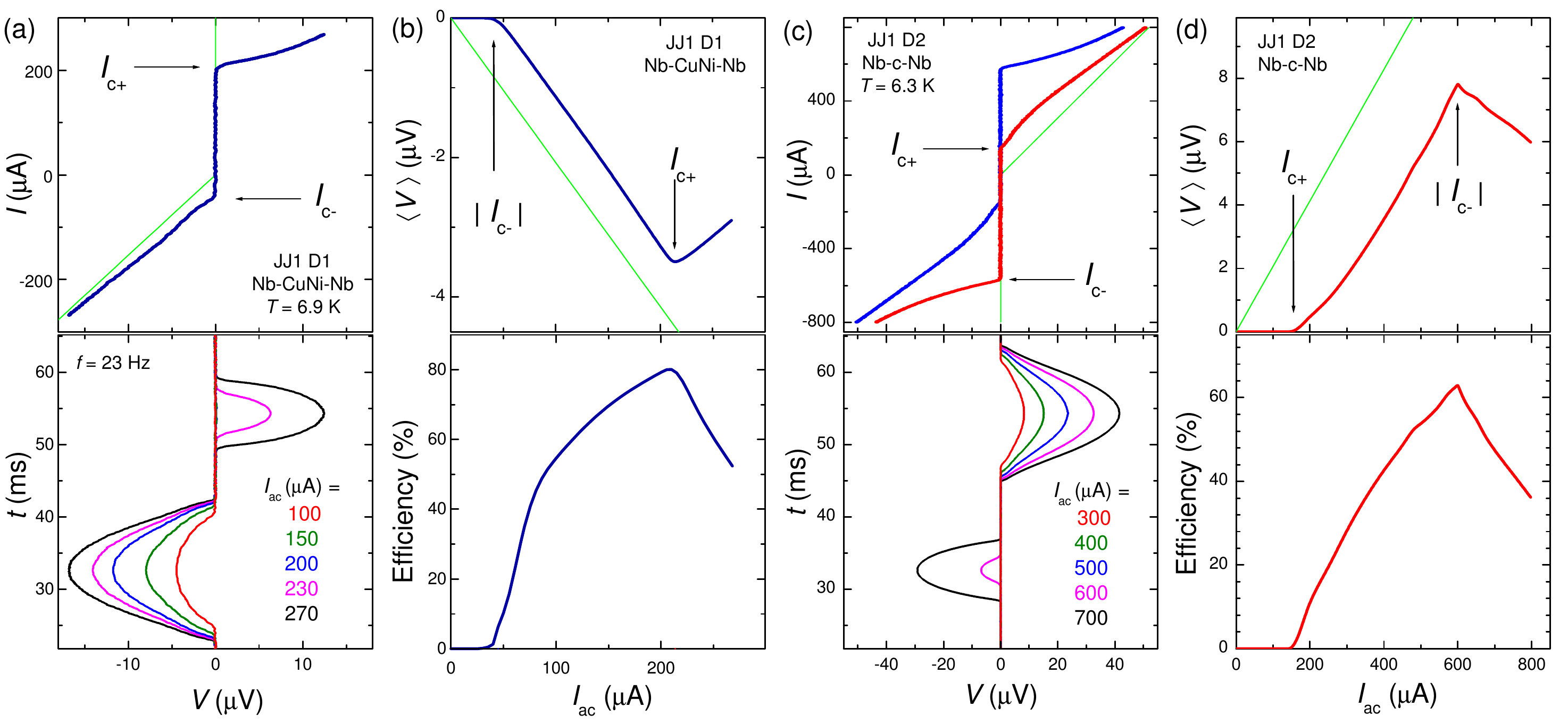}
    \caption{ {\bf Amplitude dependence of rectification.} (a) Top: the $I$-$V$ of JJ1 on D1 (dark blue) with nearly maximum nonreciprocity, $I_{c+}/|I_{c-}|$. It is measured using right-corner bias, $I=I_{ac}\sin(2\pi f t)$. Green line represents the ideal case with infinite nonreciprocity, $I_{c-}=0$. Bottom: Time dependencies of voltages during one ac-period at $f=23$ Hz for different bias amplitudes. (b) Top: time-averaged dc-voltage as a function of ac-bias amplitude for JJ1 on D1 at $T\simeq 6.9$ K. Green line represents the ideal case, $\langle V \rangle_{max}=I_{ac}R_n/\pi$. Bottom: rectification efficiency,  $\langle V \rangle/\langle V \rangle_{max}$, versus ac-bias amplitude. (c) and (d) show similar data for JJ1 on D2 at $T=6.3$ K. Top panel in (c) shows $I$-$V$s with maximum nonreciprocities obtained at $H=0.7$ Oe (red) and -0.7 Oe (blue) without AV. (d) and bottom panel in (c) represent analysis of rectification for the red $I$-$V$ with $I_{c+}<|I_{c-}|$, leading to a positive rectified voltage.
    }
    \label{fig:fig5}
\end{figure*}

Rectification is an important property of a diode. Bottom panels in Fig. \ref{fig:fig4} show rectified time-average dc-voltage, calculated for ac-bias with the amplitude $I_{ac}=220~\mu$A. Oscillatory field dependence with a significant rectified voltage at central peaks can be seen \cite{Krasnov_1997,Supplem}. The maximum rectifiable voltage for the case when one side of the $I$-$V$ is fully open, $I_{ac}<I_c$, and the other is fully closed, $I_c=0$, is $\langle V_{max} \rangle = I_{ac} R_n/\pi$. 
Central peaks for simulated red curves, which have nonreciprocities in the range of 30-50, are practically ideal. Experimental peaks for the vortex-free case (a) are exceeding $80~\%$ of that value for the two central peaks. The peaks at $H=0$ in (b) and (c) exceed $70~\%$, indicating good rectification efficiency of the diodes.

Figure \ref{fig:fig5} illustrates ac-bias dependence of rectification for D1 (a,b) and D2 (c,d). For clarty we consider the states with maximum nonreciprocity at finite $H$. The top panel in Fig. \ref{fig:fig5} (a) shows the $I$-$V$ (royal) of JJ1 on D1 with near maximum $|I_{c+}/I_{c-}|$. Bottom panel shows time dependencies of voltage, for different ac-bias amplitudes. It is seen that for $|I_{c-}|<I_{ac}<I_{c+}$ only negative voltage appears during the ac-oscillation period, leading to appearance of a negative time-average dc-voltage, $\langle V \rangle<0$. Top panel in Fig. \ref{fig:fig5} (b) shows bias dependence of rectified voltage. It appears at $I_{ac}>|I_{c-}|$, grows linearly up to $I_{ac}=I_{c+}$ and then decreases due to progressive increase of positive voltages during the oscillation period, as seen from magenta and black $V(t)$ curves in Fig. \ref{fig:fig5} (a). The green line represents the ideal case with infinite nonreciprocity, $\langle V_{max} \rangle = I_{ac} R_n/\pi$. Bottom panel in Fig. \ref{fig:fig5} (b) shows rectification efficiency with respect to the ideal case, $\langle V \rangle / \langle V_{max} \rangle $. It is seen that the maximum efficiency, achieved at $I_{ac}=I_{c+}$, exceeds $80~\%$. The maximum rectification efficiency is close to $1-1/\nu$, where $\nu$ is the nonreciprocity of $I_c$.

Figs. \ref{fig:fig5} (c) and (d) demonstrate similar data for D2. Here we analyze the state, represented by the red $I$-$V$ with maximum nonreciprocity of $|I_{c-}/I_{c+}|$. This leads to the opposite diode polarity, compared to Figs. \ref{fig:fig5} (a,c), with $\langle V \rangle >0$. The overall performance is similar to D1, except for the larger $I_c R_n$ of Nb-c-Nb JJs, which results in proportionally larger rectified voltage and upper frequency range, $\sim I_cR_n/\Phi_0$  (see Supplementary \cite{Supplem}  for additional clarifications).

To conclude, we demonstrated operation of a vortex-based Josephson diode with a large and switchable nonreciprocity at zero magnetic field.  Our concept is based on utilization of nonuniform bias for 
inducing nonreciprocity and stray fields of Abrikosov vortex,
trapped at a proper position, for shifting nonreciprocity to zero
field. It is shown that such diodes have very good performance.
Measured nonreciprocity of critical current exceeds a factor 4 at
zero field and is more that an order of magnitude at finite field.
Numerical modeling indicates that these values can be improved by
another order of magnitude by careful design. The rectification
efficiency exceeds $70~\%$ at zero field. This is good enough for
realization of more complex logical Boolean devices, needed for a
digital superconducting computer. It has already been demonstrated
\cite{Golod_2015} that a very simple geometry of such devices,
which do not utilize SQUIDs, along with a nano-scale vortex size
allows drastic miniaturization down to submicron dimensions.
However, the most unique feature of our diodes is their
switchability and tunability: (i) nonreciprocity at $H=0$ can be
easily introduced/removed by trapping/removing Abrikosov vortices
using short current pulses and (ii) diode polarity can be flipped
by changing either the vortex sign, or the bias configuration. We
argue that this may facilitate in-memory operation: an emerging
new concept capable of boosting computer performance by avoiding
bottlenecks associated with data shuffling between processor and
memory \cite{In_memory}. This could open new perspectives for
development of a digital superconducting computer.
From this perspective it is advantageous that the diode is realized 
using conventional Nb-technology, which is mature enough for large-scale applications \cite{Tolpygo_2019}.


{\bf Methods}

{\bf Samples.}

Studied devices contain planar JJs. D1 is made from  Nb (70 nm, top)/CuNi(50 nm, bottom) bilayer with superparamagnetic CuNi. D2 is made from a single Nb (70 nm) film. Films are deposited by dc-magnetron sputtering. They are first patterned into $\sim 6~ \mu$m-wide bridges by photolithography and reactive ion etching, and subsequently nano-patterned by Ga$^+$ focused ion beam (FIB). Both Nb-CuNi-Nb (D1) and Nb-c-Nb (D2) JJs have a variable-thickness-bridge structure. They are made by cutting a narrow (20-30 nm) groove in the top Nb layer by FIB. Vortex trap (a hole $\sim 50$ nm in diameter) is also made by FIB. 
Both devices have similar cross-like geometry, as can be seen from Fig. \ref{fig:fig3} (b) and Supplementary Figure 1. They have practically identical dimensions, specified in details in the Supplementary \cite{Supplem}. We fabricated and tested similar JJs with other metals in the bottom layer  \cite{Golod_2010,Golod_2015,KrasnovSFS_2005,Golod_2019a,Golod_2019b,Boris_2013,Grebenchuk_2022}. All of them work in a similar manner and results do not depend on the presence or specific material of the bottom layer.

{\bf Experimental details.}

Measurements were performed in a closed-cycle cryostat. Magnetic field is applied perpendicular to the film (positive $H$ along $y$-direction). $I_c(H)$ patterns were automatically recorded upon sweeping of magnetic field, provided by a superconducting solenoid. For this current-voltage characteristics were examined and $I_c$ was determined using a small threshold voltage criterion, $V_{th}\sim 1-2~\mu$V. All $I$-$V$s shown in the manuscript are nonhysteretic.

{\bf Numerical simulations.}
Critical current is calculated by maximization of the Josephson current,
\begin{equation}
I_s = \int_0^L{J_c(x) \sin [\varphi(x)+\varphi_0] dx}
\label{Is},
\end{equation}
with respect to the phase offset $\varphi_0$. Here $J_c(x)$ is the critical current density along the JJ. The Josephson phase difference $\varphi(x)$ has two contributions \cite{Krasnov_2020}:
\begin{equation}
\varphi(x)=\frac{2\pi d_{eff}}{\Phi_0} B_y x + \varphi_v(x),
\label{fx}
\end{equation}
where $d_{eff}$ is the magnetic thickness of the
junction. Here the first term represents the linear phase gradient induced by the $y$-component of magnetic induction and the second - a nonuniform phase shift induced by the trapped Abrikosov vortex. Since we consider only short junctions, we neglect possible screening effects and assume that $B_y$ is uniform ($x$-independent). However, to account for the self-field effect we add an extra contribution to the applied external field $H$,
\begin{equation}
B_y = H + L_{sf} I_s,
\label{By}
\end{equation}
proportional to the total current and the self-field inductance, $L_{sf}$. The vortex contribution is given by the azimuthal angle \cite{Golod_2019b}, indicated in the sketch in Fig. \ref{fig:fig1} (b),
\begin{equation}
\varphi_v(x)= -V \arctan \left(\frac{x-x_v}{|z_v|}\right),
\label{fv}
\end{equation}
where $V$ is the vorticity. Due to the self-field term in Eq. (\ref{By}), $I_s$ is present in the right-hand-side of Eq. (\ref{Is}) as well.
This implicit equation is solved iteratively using the bisection method. The fitting is obtained by varying two constants: $L_{sf}$ in Eq. (\ref{By}) and $V$ in Eq. (\ref{fv}); as well as allowing for a nonuniform $J_c(x)$ distribution in Eq, (\ref{Is}). The fit represented by magenta line in Fig.\ref{fig:fig1} (a) and red lines in Fig. \ref{fig:fig4} corresponds to the V-shaped $J_c(x)$ with a $25\%$ $J_c$ reduction in the middle of the JJ, $x=L/2$.


\end{document}